# 3D-Hydrogen Analysis of Ferromagnetic Microstructures in Proton Irradiated Graphite

P. Reichart[1], D. Spemann[2], A. Hauptner[3], A. Bergmaier[4], V. Hable[4], R. Hertenberger[5], C. Greubel[4], A. Setzer[2], G. Dollinger[4], D.N. Jamieson[1], T. Butz[2], P. Esquinazi[2]

[1]*Microanalytical Research Centre, School of Physics, University of Melbourne, Parkville 3010 VIC, Australia*
[2]*Institut für Experimentelle Physik II, Universität Leipzig, 04103 Leipzig, Germany*
[3]*Physik Department E12, Technische Universität München, 85748, Garching, Germany*
[4]*Institut für Angewandte Physik und Messtechnik, Universität der Bundeswehr München, 85577 Neubiberg, Germany*
[5]*Department für Physik, Ludwig-Maximilians-Universität München, 85748 Garching, Germany*

Abstract

Recently, magnetic order in highly oriented pyrolytic graphite (HOPG) induced by proton broad- and microbeam irradiation was discovered. Theoretical models propose that hydrogen could play a major role in the magnetism mechanism. We analysed the hydrogen distribution of pristine as well as irradiated HOPG samples, which were implanted to µm-sized spots as well as extended areas with various doses of 2.25 MeV protons at the Leipzig microprobe LIPSION. For this we used the sensitive 3D hydrogen microscopy system at the Munich microprobe SNAKE. The background hydrogen level in pristine HOPG is determined to be less than 0.3 at-ppm. About $4.8 \times 10^{15}$ H-atoms/cm$^2$ are observed in the near-surface region (4 µm depth resolution). The depth profiles of the implants show hydrogen located within a confined peak at the end of range, in agreement with SRIM Monte Carlo simulations, and no evidence of diffusion broadening along the c-axis. At sample with microspots, up to 40 at-% of the implanted hydrogen is not detected, providing support for lateral hydrogen diffusion.

Corresponding Author: Patrick Reichart, Microanalytical Research Centre, School of Physics, University of Melbourne, Parkville 3010 VIC, Australia. Email: patrick.reichart@unimelb.edu.au, Fax: +61 3 93474783

1. **Introduction**

Several reports of ferromagnetic loops in highly oriented pyrolytic graphite (HOPG) [1] or polymerised fullerene [2] have shown that ferromagnetism in carbon-based structures containing only p- and s-electrons is possible and furthermore stable at room temperature [2,3]. Recent studies on HOPG have ruled out the possibility that magnetic impurities are the origin [4,5] and provide striking evidence that ferromagnetism can be induced reproducibly in HOPG by MeV proton irradiation [3,6-8]. The measurements indicate a highly localized magnetic moment in the order of $10^5$ $Am^2$ [7], although the exact quantification of the magnetization is difficult as the spatial extent of the ferromagnetic region is not clear.

Theoretical models suggest that the magnetism is due to, for example, a mixture of $sp^2$-/$sp^3$-carbon atoms with an unpaired π-electron ferromagnetically aligned [9] or spontaneous magnetization due to different numbers of mono- and dihydrogenated carbon atoms [10]. In order to test whether hydrogen plays a role the precise knowledge of the hydrogen distribution in the HOPG samples is necessary. However, deeply implanted hydrogen is very difficult to analyse and, in particular, when it comes to micrometer dimensions, the sensitivity of standard methods like Magnetic Resonance Imaging, Nuclear Reaction Analysis or Elastic Recoil Detection is limited to values far above the ppm region. The common nuclear microprobe analysis methods suffer mainly from irradiation damage [11]. One unique alternative exists with the new 3D hydrogen microscopy system at the Munich microprobe SNAKE [12] which uses coincident proton-proton scattering [13] combined with a scanning 17 MeV proton microbeam. This method has sufficient sensitivity below 0.1 ppm for analysis within micrometer resolution, as recently demonstrated by the imaging of hydrogen on grain boundaries of diamond [14]. With this, we were able to address three basic questions:

(1) What is the hydrogen content in the pristine HOPG samples? (2) Is the implanted

hydrogen located at the implanted sites or is a spread laterally/in depth observed? (3) How much hydrogen diffuses out of the implanted microspots, depending on the implanted dose?

After the description of the sample preparation, the proton implantation and the experimental setup, we report on our measured hydrogen distribution data with detailed discussions of the depth profiles and lateral distributions.

## 2. Experimental

### 2.1. Sample preparation

The samples for the present study were implanted with 2.25 MeV protons on the microprobe LIPSION parallel to the c-axis like for the earlier ferromagnetism studies [7], using the same material (ZYA grade HOPG, Advanced Ceramics Co.). For this purpose, the samples were glued with varnish over 5 mm apertures which are drilled into the aluminium sample holders dedicated for transmission analysis at the SNAKE microprobe setup. Then the HOPG samples were thinned in order to get freestanding sheets of HOPG where the protons are able to pass through under all scattering angles from 30° – 60° for the coincidence analysis.

Three samples have been prepared for the hydrogen analysis study:

(A) 25 μm thick pristine HOPG to measure the intrinsic H content.

(B) 120 μm thick HOPG with four implanted areas. Each area was implanted with a broad beam through an aperture with a charge of 160 μC at an area of approximately 1 mm$^2$, resulting in a fluence of about $1 \times 10^{17}$ cm$^{-2}$ (ion flux $\approx 10^{14}$ cm$^{-2}$s$^{-1}$).

(C) 120 μm thick HOPG with 2 × 2 rows each with 10 implanted microspots separated by 20 μm. The implanted charge was 85, 34, 17, 8.5, 3.4, … , 0.085 nC for each row, corresponding to (5.3, 2.1, 1.1, … , 0.0053) × 10$^{11}$ protons. The implanting focused beam had a Gaussian profile with about 1.8 μm FWHM and a current of 330 pA, which results in an ion flux of about 10$^{17}$ cm$^{-2}$s$^{-1}$. Clear magnetic force microscopy signals have been observed previously for such conditions, which correspond to a proton fluence of about $2\times10^{19}$ cm$^{-2}$ down to $2\times10^{16}$ cm$^{-2}$, if a homogeneous spot area of this FWHM diameter is assumed.

*2.2.    Setup of 3D hydrogen analysis*

The 3D hydrogen microscopy setup at the microprobe SNAKE located in the Munich 14 MV tandem accelerator lab is described in detail in earlier publications [15]. The central part of the coincident proton detection system is a large annular silicon PIN diode divided into 16 sectors and 48 rings combined with a multi channel readout and fast coincident timing analysis. The detector covers scattering angles between 30° and 60° in transmission geometry behind the sample, resulting in a solid angle of detection of 2.3 sr. This gives an optimum benefit regarding the lowest damage potential of the method and optimizes the time needed to acquire scan maps at low hydrogen contents. The incident proton energy with this 1 mm thick detector is limited to 17 MeV which is nevertheless sufficient to analyse freestanding graphite samples up to 150 μm thickness. At these proton energies, hydrogen depth profiling is possible up to a depth resolution of about 3 μm by correcting for the path length effect of the scattered protons [15]. The depth scale was calculated from the difference between the sum of energy losses of the two coincident protons and that of the incoming proton with the stopping power data of Ziegler et al. [16]. We used a HOPG density of 2.26 g/cm$^3$ and the cross section data of Burkig et al. [17] for the determination of the hydrogen depth profiles.

The system was calibrated using a 0.9 μm thick aluminized Mylar foil, 130 μm Kapton and Polycarbonate sheets (60 μm and 250 μm) as hydrogen standards. The nominal composition of these hydrogen standards has been verified using heavy ion elastic recoil detection and the results were consistent with the chemical formula for these foils. With this calibration, systematic errors in the charge integration and detection efficiency are determined with an accuracy of about 10%. The latter is mainly affected by losses in the coincidence readout electronics and software coincidence filters. Additional losses of true coincidence events arise at large depth due to enhanced lateral straggling along the outgoing paths inside the sample. These losses have been taken into account in case of the thick graphite samples by adopting the depth dependent efficiency from the thick hydrogen standards.

For analysis each graphite sheet is placed in the focal plane of the microprobe. The beam was focused to a beam spot of 1.5 μm diameter which represents the lateral resolution of the hydrogen microscopy for this experiment.

## 3. Results and discussion

### 3.1. Pristine HOPG

The pristine HOPG sample (A) was investigated on a scan area of $200 \times 200$ μm$^2$ with an average proton current of about 100 pA and a total applied fluence of $2.7 \times 10^{16}$ cm$^{-2}$ (0.04 nC/μm$^2$).

Fig. 1 shows the hydrogen depth profile with a depth resolution of about 4 μm. Hydrogen on the surfaces appears as two peaks with this FWHM. The hydrogen scan map (not shown) reveals that this hydrogen is homogenously distributed over the scan area. The hydrogen yield corresponds to 7.0 and $4.8 \times 10^{15}$ H-atoms/cm$^2$ on the front and back surface, respectively. As there was no special surface treatment, we conclude that this represents the usual surface termination with hydrocarbon and water molecules.

This surface hydrogen coverage is of the same order of magnitude as the minimum proton fluence which leads to the observed formation of ferromagnetism. Thus far, this surface contamination has not been considered to influence the formation of ferromagnetic regions although it might not be neglected in graphite which is damaged by proton as well as non-hydrogen irradiation.

The background in the depth profile of Fig. 1 is caused by accidental coincidences. Events at unphysical energies or depths left of the left peak and right of the right peak, i.e. in vacuum, represent this background level corresponding to a hydrogen content of $< 5.6\times10^{16}$ atoms/cm$^3$ or 0.5 at-ppm. If this background is subtracted and statistical errors are considered, the hydrogen content in the bulk region between 5 – 18 µm is zero but consistent with an upper limit of $3.3\times10^{16}$ atoms/cm$^3$ or 0.3 at-ppm.

*3.2.   Large area implants (B)*

The HOPG sample (B) with the four large implanted spots was analysed with a microbeam scanned over an area of $200 \times 200$ µm$^2$ near the centre of one spot with an average beam current of 21 pA. The scan map (not shown) confirms again a homogeneous distribution of hydrogen at the surfaces as well as in the implanted bulk region. However, on the surface opposite to the implanted side we detected a hydrogen contamination of $4.5\times10^{18}$ cm$^{-2}$, which is probably caused by glue residues. This surface coverage is more than two orders of magnitudes larger than at the opposite surface which exhibits $1.4\times10^{16}$ cm$^{-2}$.

Fig. 2 shows the corresponding depth profile with the glue peak on the left (region I), the implanted peak (III) and the right surface peak (V). The detected hydrogen is located at the end of range at 46 µm in agreement with Monte Carlo simulations [16] (SRIM, v.2003.26, HOPG density 2.26 g/cm$^3$) and the peak shows no significant tails above the background. Within the depth resolution of the method, the implantation peak has not been broadened compared to the simulated implantation profile suggesting that no diffusion has occurred along the HOPG c-axis. Making up the hydrogen balance, the content in the implantation peak corresponds to $1.1\times10^{17}$ cm$^{-2}$. This value is apparently in agreement with the nominal implanted hydrogen. However, we note here an uncertainty of about 50% for the implanted fluence as the small scan size allows no proof of the homogeneity as well as the actual size of the implanted area.

The hydrogen content detected in the regions II (15 μm – 60 μm) and IV (80 μm – 105 μm) gives a background corrected value of (2.5 ± 0.9) at-ppm and (3.0 ± 1.0) at-ppm, respectively. No pile-up correction is required because of suppression by the tight coincidence filters, and moreover, the peaks III and V show no tails. Therefore, this measured H content is significantly above the pristine bulk hydrogen content. The background content integrated across the whole HOPG thickness amounts to $3.3 \times 10^{15}$ cm$^{-2}$, which equals 3% of the implanted fluence.

*3.3. Implanted microspots (C)*

The sample with the implanted microspots was analysed with a microbeam scanned at the spots with the five largest implanted doses. These implanted spots are clearly visible in the optical micrograph of Fig. 3a due to swelling and the outlined rectangle marks the analysed area of $20 \times 86$ μm$^2$ (analysing current I = 13 pA, integrated charge Q = 130 nC).

The depth profile shown in Fig. 3b shows on the right side at maximum depth the peak due to the glue residue contamination. Note that this profile is reversed compared to Fig. 2 because a reversed analysing direction was required to allow for optical positioning of the microspots. The drawback of this is clearly a significant background of hydrogen events due to a tail from the large H peak contributed by the glue residues. This tail is more likely caused by coincidence events with a missing fraction of the sum energy due to detector effects rather than by hydrogen located at this depth. The hydrogen scan map (Fig. 3e) with hydrogen events originating from this surface region III as marked in Fig. 3b reveals that the glue residue contamination is localized at the top area containing the fifth implanted spot only. Therefore, the background level at this spot with nominally $2 \times 10^{10}$ implanted protons is very close to the "true" hydrogen signal but prevents the analysis of further spots with still lower fluence further up on the sample. The background due to accidental coincidences ranges from (4.4 ± 0.4) at-ppm at z = 0 to (1.0 ± 0.2) at-ppm at z = 130 μm. As determined from the depth profile with hydrogen events originating from the region excluding the contaminated scan area (Fig. 3f), the background corrected hydrogen level in the bulk regions left and right of region II corresponds to (1.6 ± 0.5) at-ppm to (3.2 ± 0.6) at-ppm, respectively. The latter has a significant contribution from a small peak around z =85 µm. These events originate from an area located at the lower left corner of the scan region (x < -4 µm, y < 10 µm) only and are not correlated to the spot positions.

Therefore, the sensitivity in the major area of the scan is easily sufficient to resolve the implanted hydrogen. This is shown in the scan map in Fig. 3d with hydrogen events originating from the marked implantation region II only, i.e. a depth of 40 to 70 µm. The map shows spots with a Gaussian distribution of the events with a FWHM (= 2.35σ) in the x-position ranging from 5.5 µm to 3.9 µm (y: 5.2 µm – 2.8 µm). The widths are well in agreement with the geometric sum of the contributions of the implanting beam width (1.8 µm), the implanting radial spread (~ 5 µm[1]), the analysing beam (1.5 µm) and a negligible analysing beam spread. Furthermore, the background level (accidental coincidences and bulk hydrogen content) is reached within a radius of 3σ around the centre of the spots. This proves that the hydrogen is well located at the implanted sites giving no evidence of a diffusion broadening on a micrometer scale.

However, making up the hydrogen balance with the background subtracted as shown for the spots in Tab. 1, we observed that 18 – 36% of the implanted hydrogen is not detected in the microspots within 3σ distance around the centre and this loss is not correlated to the implanted fluence. The accuracy of the given implanted charge value is better than 10%, based on inaccuracies in the non-Rutherford cross sections and statistics, and the accuracy of the hydrogen calibration was noted as 10% only as well. Therefore this loss of hydrogen is significant and loss of hydrogen during the proton-proton scattering analysis due to irradiation damage can be excluded as no decrease of the hydrogen yield is observed as shown in Fig. 4.

---

[1] It was taken into account that SRIM underestimates lateral straggling for protons up to 45% for Mylar [21] which is assumed to be similar for graphite. This discrepancy still exists in the latest version of SRIM (2003.26).

The most likely explanation might be diffusion or trapping/detrapping over a distance larger than the scan area or macroscopic distances, in particular along the orientated planes in HOPG. In isotropic graphite, models for hydrogen trapping [18] suggest fast molecular diffusion on crystallite surfaces or graphite planes and recent absorption/desorption experiments [19] report a low activation energy for H detrapping from edge surfaces. However, previous investigations by Siegele et al. [20] indicate 100% trapping of low energy implanted deuterium in HOPG over a large fluence range of $10^{15}$ up to $7.5 \times 10^{17}$ D/cm$^2$ at room temperature. We note here that diversity in the used HOPG material has carefully to be taken into account when comparing these results. The fluence in our microspots reaches up to $10^{19}$ H/cm$^2$. Additionally, the high flux (ions×cm$^{-2}$×s$^{-1}$) of microprobe implantation causes a power density of several 100 W/mm$^2$ in the implanted spots and locally elevated temperatures could cause indeed a release of a larger fraction of the implanted hydrogen into undamaged regions and diffusion out of the scan area. Thermal effects have also been observed in recent studies regarding the magnetic order in proton irradiated HOPG [22,23]. Whether the observed hydrogen loss is indeed a thermal effect, i.e. flux dependent has to be investigated in future experiments.

Regarding the surface hydrogen, Fig. 3c shows the hydrogen map of the implanted "front" surface region I. It reveals a homogeneous hydrogen distribution and it shows no evidence of enhanced trapping of hydrogen at the damaged near surface areas within the sensitivity of the method.

## 4. Conclusion

The 3D hydrogen microscopy system at SNAKE has been shown to have sufficient sensitivity and spatial resolution to investigate the behaviour of micro implanted hydrogen as it is used in studies of magnetic carbon. It is therefore so far the only option for hydrogen analyses at these low hydrogen concentrations. We determined an upper limit of 0.3 at-ppm of hydrogen in the bulk of the pristine HOPG sample. The observed hydrogen on the surface has not yet been considered as relevant to the formation of magnetic order in graphite. However, the detected amount of up to $10^{16}$ H-atoms/cm$^2$ on the untreated surfaces might play a major role if incorporated properly in the appropriate sites of graphite after ion irradiation. This working hypothesis has to be investigated both experimentally and theoretically in continued studies.

Furthermore, we confirmed that the implanted hydrogen stays well located at the implanted sites, both laterally as well as in depth at the end of range. This is true even for microspots with highest hydrogen fluence where a peak hydrogen content of about 15 at-% is measured. At the implanted microspots, we discovered a significant loss of hydrogen for all analysed spots, although the lateral profile is in agreement with the calculated range and damage distribution and there is no evidence of a diffusion broadening. This suggests that a part of hydrogen is released from implanted sites, possibly due to thermal effects at high beam fluxes, and can thus diffuse laterally over larger distances. Part of this hydrogen is found quite homogeneously distributed within the whole graphite thickness as shown in the experiment of the large area implant.

Acknowledgements

This work was supported by the Deutsche Forschungsgemeinschaft under grant DFG Es86/11-1.

Table 1:

| spot no. | detected hydrogen events within r = 3σ | hydrogen content (× $10^9$ atoms) | | |
|---|---|---|---|---|
| | | measured (background subtracted) | nominally implanted | discrepancy |
| 1 | 872 | 406 | 550 | 26% |
| 2 | 316 | 147 | 220 | 33% |
| 3 | 150 | 70 | 110 | 36% |
| 4 | 90 | 42 | 55 | 24% |
| (5) | (38) | (18) | (22) | (18%) |

Table 1: Measured and nominally implanted hydrogen content of the implanted spots within a radius of 3σ = 7 μm around the centre of each spot. The background was determined from layers next to the implanted one (glue residue area excluded) and corresponds to 11.4 hydrogen events normalized to the 3σ area.

Figure 1:

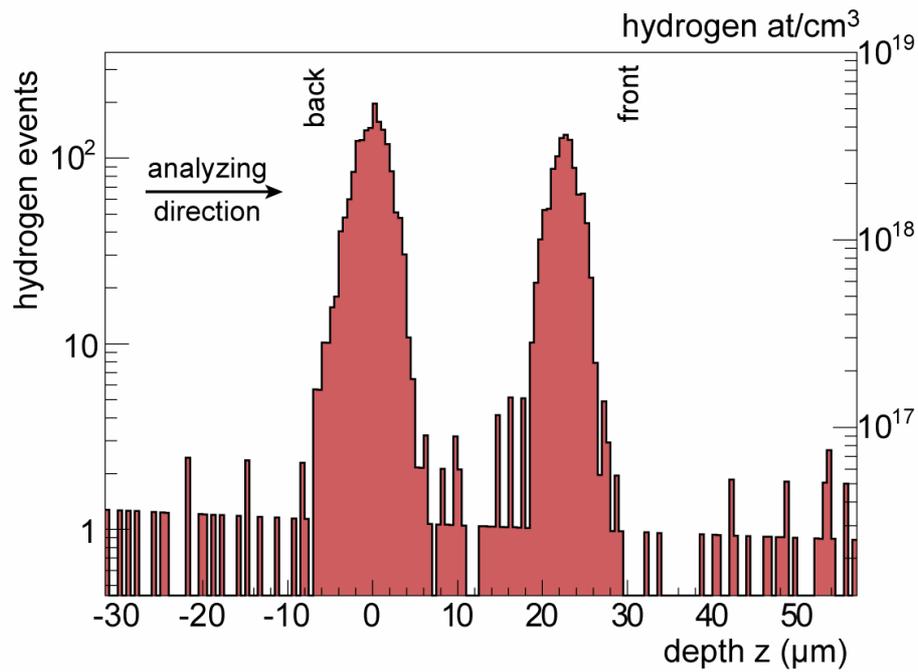

Figure 1: Hydrogen depth profile of the pristine HOPG sample A, obtained from a 200 × 200 μm² area (~ 4 μm depth resolution). The two peaks represent the surface contamination with 7.0 and $4.8\times10^{15}$ H-atoms/cm² for the back and front peak due to surface adsorbates, respectively. The hydrogen yield is calibrated with an efficiency function depending on the depth of the coincidence events. This scales also the background caused by accidental coincidences. The hydrogen content between the peaks is consistent with an upper limit of 0.3 at-ppm.

Figure 2:

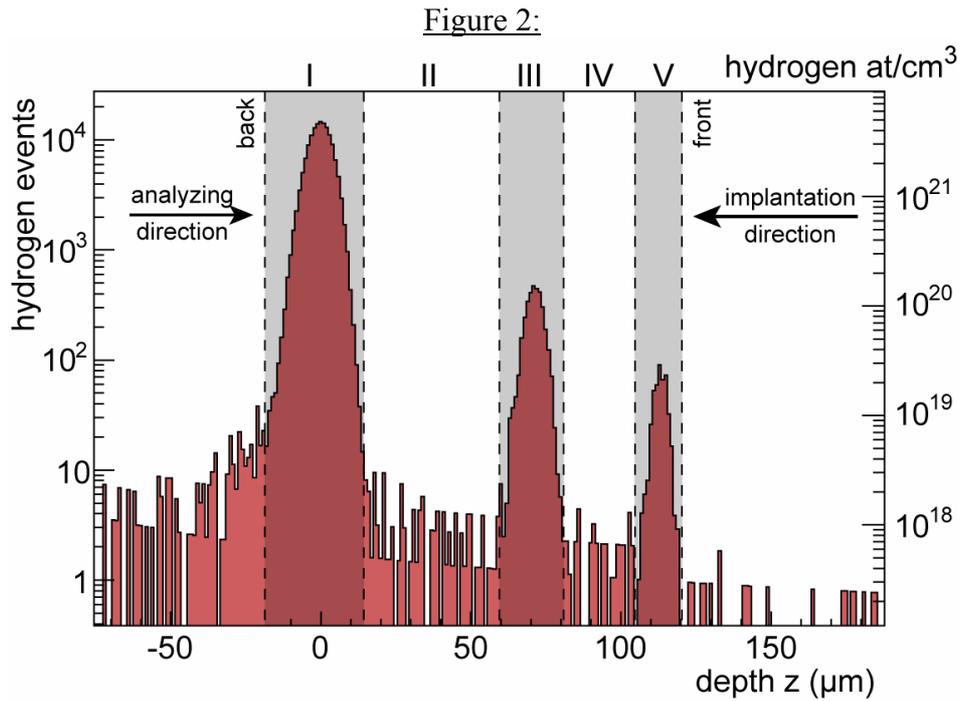

Figure 2: Hydrogen depth profile of the HOPG sample B (large implanted area). The middle peak (region III) is located at the end of range of implanted 2.25 MeV protons and the peak content correspond to $1.1 \times 10^{17}$ cm$^{-2}$. The left and right peaks represent the hydrogen contamination on the surfaces (glue residue contamination on the left "back" surface). The depth resolution of the analysis ranges from 9 μm at z = 0 to 4.7 μm at the right peak.

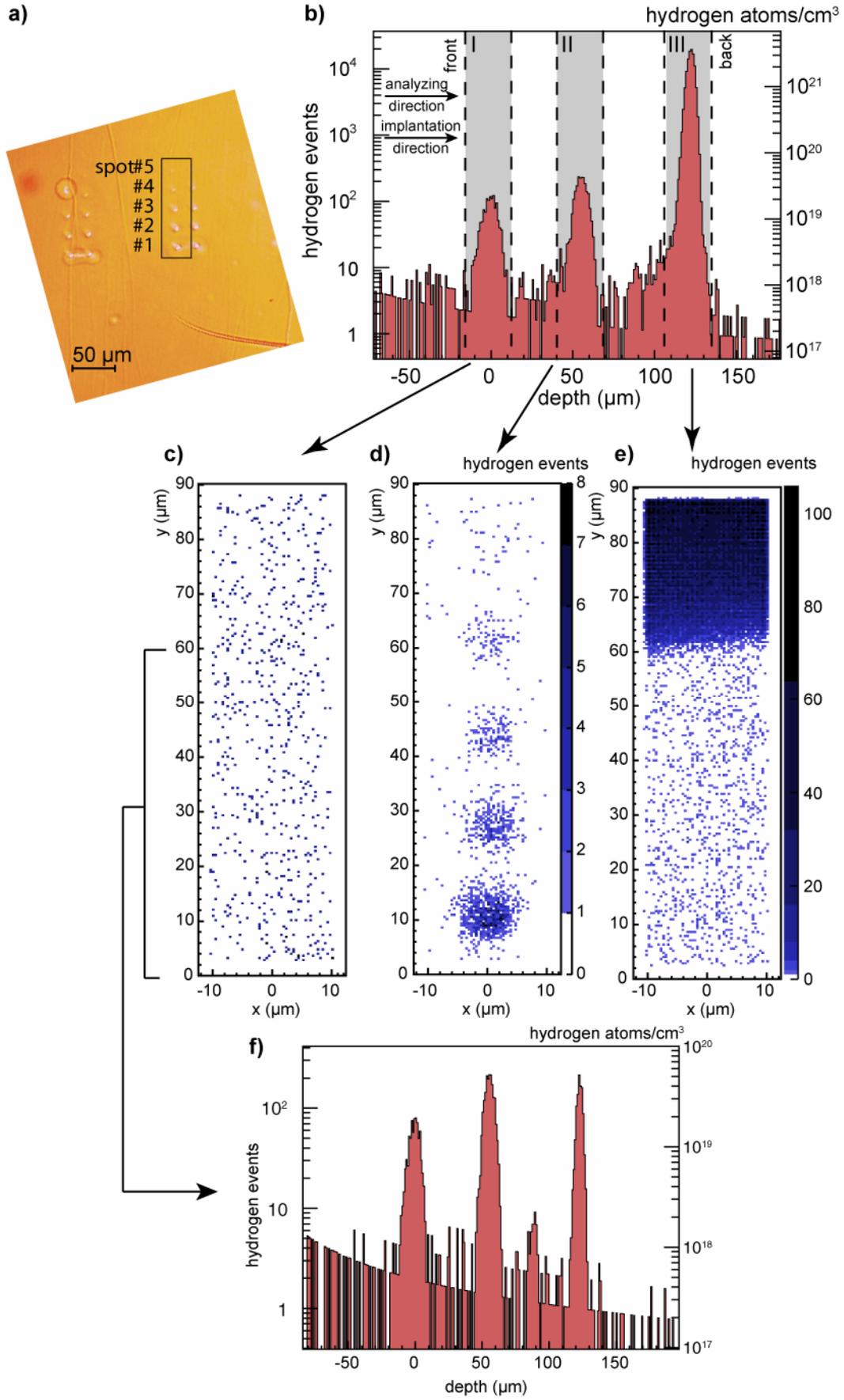

Figure 3:

Figure 3: (a) Optical image of the implanted spots on sample C. (b) Hydrogen depth profile of the scanned region marked by the black rectangle in Fig. 3a with the peak from the "front" surface (region I), the implanted peak (region II) and on the right the large hydrogen peak due to glue residue contamination from the "back" surface (region III). (c) Scan map of hydrogen events originating from the surface region I. One entry in the map corresponds to about $3.4\times10^8$ hydrogen atoms. (d) Scan map of hydrogen events originating from the region II (implanted hydrogen). (e) Hydrogen map of surface region III showing elevated levels around the position of spot #5 only (due to glue residue at the back). (f) Depth profile with the glue contaminated area excluded, i.e. for y-positions 0 – 60 μm.

Figure 4:

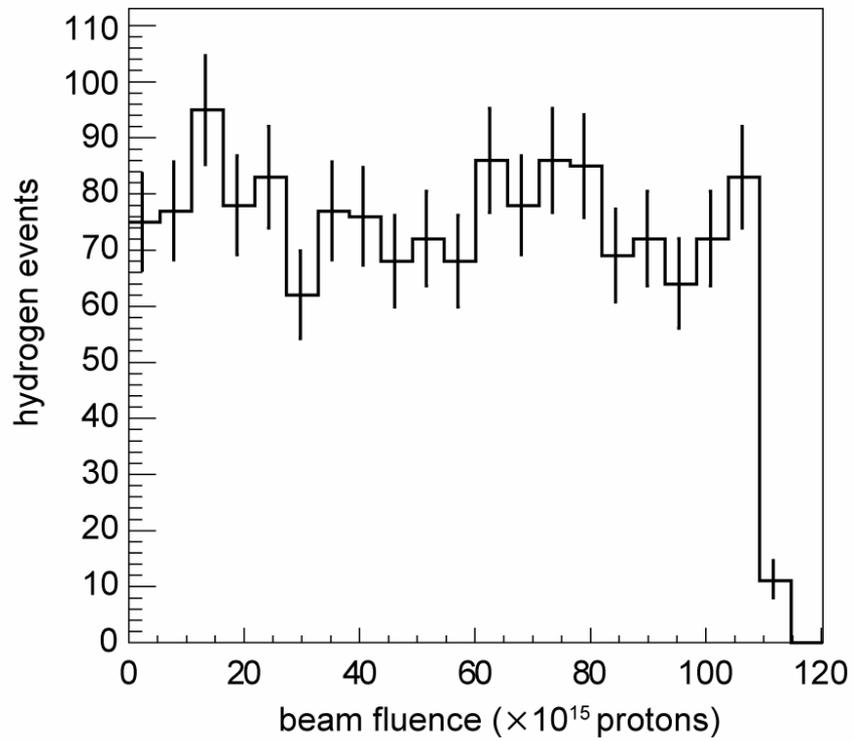

Figure 4: Hydrogen yield originating from the spots vs. the applied analysing beam fluence. No significant loss of hydrogen due to irradiation damage is observed during the hydrogen analysis.